\documentclass[12pt]{iopart}
\usepackage{iopams}
\usepackage{geometry}
\usepackage{graphicx,picture}
\usepackage{cite}

\begin{document}

\eqnobysec
\title[]{ Thermal R-current correlators from AdS/CFT correspondence}

\author{Xuanmin Cao$^1$, Lian Liu$^1$ and Hui Liu$^1$}

\address{$^1$ Department of Physics, Jinan University, Guangzhou (510632) China}

\ead{tliuhui@jnu.edu.cn}

\begin{abstract}
 We calculate all components of thermal R-current correlators from AdS/CFT correspondence for non-zero momentum and energy. In zero momentum limit, we find an analytic expression for the components $G_{xx}(G_{yy})$. The dielectric function of strong coupling is also presented and compared with that in weak coupling.

\end{abstract}
\pacs{11.10.Wx, 11.25.Hf, 12.38.Mh, 82.20.Sb}
\maketitle

\section{Introduction}

In heavy ion collision experiments, more and more evidences support the fact that the quark gluon plasma (QGP) produced in Au-Au collision is strongly coupled rather than weakly coupled as we expected \cite{a,b,c,d,e,f}.  However strong coupling is a real challenge in theory since perturbative field theory fails when the coupling constant is larger than one. The AdS/CFT correspondence \cite{g,h,i,j,k}, which shows the relation between string theory on AdS spacetime with conformal field theory on boundary of the bulk spacetime, has emerged as an extremely powerful tool for dealing with strong coupling QGP problems. The simple but still useful example of such duality is the Maldacena conjecture \cite{g}: the weak coupling limit in type IIB string theory on $\textrm{AdS}_5\times \textrm{S}^5$ space is equivalent to the strong coupling limit in four-dimensional $\mathcal{N}=4$ supersymmetric Yang-Mills(SYM) theory. The conjecture makes it possible to understand the field-theory correlation functions in the strong coupling through weakly coupled classical supergravity.

Although extracting the Lorentzian-signature AdS/CFT results directly from gravity is indispensable, many useful properties of gauge theories at finite temperature can be obtained from real-time Green's functions by the prescription \cite{l} of Minkowski-space correlators in AdS/CFT correspondence.  The thermal R-current correlators in $\mathcal{N}=4$ SYM were derived in the small momentum and energy limit ($\omega\ll T, q\ll T$) \cite{m} and numerically calculated at the condition of $q=0$ \cite{n}, while the correlators in full regime with respect to energy and momentum are still absent. All kinds of limit can not completely represent the full properties of correlators. So in this paper we present the thermal correlators numerically for whole range of momentum and energy. As an application, we disscuss the dielectric function.

We organize the paper as follows. In section \ref{functions}, we review the retarded correlators in the prescription of AdS/CFT correspondence. The numerical results of thermal R-current correlators are presented in section \ref{nresults}. In section \ref{dielectric}, the dielectric function is computed and compared with weak coupling. The analytic $G_{xx}(G_{yy})$ at $q=0$ are discussed in section \ref{aresults}. Section \ref{conclusions} is conclusion.

\section{Recipe for Retarded correlatos}\label{functions}
The Minkowski version of the AdS/CFT correspondence is formally written as the following equality \cite{l}.
\begin{equation}\label{correspondence}
\langle \rme^{\rmi\int_{\partial M}\phi_0\mathcal{O}}\rangle=\rme^{\rmi S_{cl}[\phi_0]},
\end{equation}
where the left hand side is the partition function of a CFT with the boundary values acting as sources for the primary operators. The partition function on the right hand side is a function of the boundary values of a quantum gravity theory on an asymptotically AdS space time.

The Retarded correlators in Minkowski space is defined as \cite{o}:
\begin{equation}
G^R_{\mu\nu}(\omega,\mathbf{q})=-\rmi\int \rmd ^4x \rme^{-\rmi q\cdot x}\theta(t)\langle [j_{\mu}(x),j_{\nu}(0)]\rangle,
\end{equation}
where $j_{\mu}$ is the charge current. The prescription that connects field theory and supergravity according to \eref{correspondence} is constructed by \cite{l},
\begin{equation}\label{pgreen}
G^R_{\mu\nu}=K_{\mu\nu}(u)f(-q,u)\partial_{u} f(q,u)|_{u\rightarrow 0},
\end{equation}
where $A_{\mu}(q,u)=f(q,u)A_{\mu}(q,u=0),f(q,0)=1$.

The background is a solution of the type IIB low energy equations of motion in the near-horizon limit $r<<R$, which is given by the metric
\begin{equation}
\fl \rmd s_{10}^2=\frac{(\pi T R)^2}{u}(-f(u)\rmd t^2+\rmd x^2+\rmd y^2+\rmd z^2)+\frac{R^2}{4u^2f(u)}\rmd u^2+R^2\rmd\Omega_5^2,
\end{equation}
where $u=r^2_0/r^2$ and $f(u)=1-u^2$, with $T=r_0/\pi R^2$  the Hawking temperature. The horizon corresponds to $u=1$ and the spatial boundary corresponds to $u=0$.

Here we consider the U(1) R-charge currents \cite{r}. The relevant  part to  correlators in the background is
\begin{equation}\label{action}
S=-\frac{N^2}{64{\pi}^2R} \int \rmd^5x \sqrt{-g}F_{\mu\nu}F^{\mu\nu},
\end{equation}
where $F_{\mu\nu}=\partial_u\mathcal{A}_{\mu}-\partial_u\mathcal{A}_{\nu}$. Under the gauge condition $\mathcal{A}_u=0$, action \eref{action} is still Lorentz-covariant, so we can get the five dimensional Maxwell equation,
\begin{equation}
\partial_{\mu}\sqrt{-g}F^{\mu\nu}=0.
\end{equation}

In order to simplify the calculation we set $q_\mu=(\omega,0,0,q)$, then the five dimensional Maxwell equation is reduced to the following set of equations:

\numparts
\begin{eqnarray}
\tilde{\omega} A^\prime_t+\tilde{q}f A_z^\prime=0\label{motiona},\\
A_t^{\prime\prime}-\frac{1}{uf}(\tilde{q}^2 A_t+\tilde{\omega}\tilde{ q}A_z)=0\label{motionb},\\
A_{z}^{\prime\prime}+\frac{f^\prime}{f}A^\prime_{z}+\frac{1}{u f^2}(\tilde{\omega}^2 A_z+\tilde{\omega}\tilde{ q} A_t)=0\label{motionc},\\
A^{\prime\prime}_{\alpha}+\frac{f^\prime}{f}A_{\alpha}^\prime+\frac{1}{uf}(\frac{\tilde{\omega}^2}{f}-\tilde{ q}^2 )A_{\alpha}=0\label{motiond},
\end{eqnarray}
\endnumparts
where $\tilde{\omega}=\omega/2\pi T,\tilde{ q}={q}/2\pi T$ are dimensionless energy and momentum respectively, $\alpha$  stands for $x$ or $y$, the prime on the field components indicate the derivatives of u. Fourier transformation has been involved to the bulk field,
\begin{equation}
\mathcal{A}_{\mu}=\int\frac{\rmd^4q}{(2\pi)^4}\rme^{-\rmi q\cdot x}A_{\mu}(q,u).
\end{equation}

Obviously the four Maxwell equations are not independent: from \eref{motiona} and \eref{motionb} we can get \eref{motionc}. Combining \eref{motiona} and \eref{motionb}, one can obtain the differential equation of $A_t$,
\begin{equation}\label{motionA}
A^{\prime\prime\prime}_t+\frac{(uf)^\prime}{uf}A_t^{\prime\prime}+\frac{\tilde{ \omega}^2-\tilde{ q}^2f}{uf^2}A_t^\prime=0.
\end{equation}

The terms in the action which contain two derivatives with respect to $u$ are
\begin{eqnarray}
S&=&-\frac{N^2}{32{\pi}^2R}\int \rmd u\rmd x^4\sqrt{-g}\ g^{\mu\nu}g^{uu}\partial_{u}\mathcal{A}_{\mu}\partial_u \mathcal{A}_{\nu}+\cdots \nonumber\\
&=&-\frac{N^2}{32{\pi}^2R}\int \rmd u\int\frac{\rmd^4q}{(2\pi)^4}\sqrt{-g}g^{\mu\nu}g^{uu}A_{\mu}^{\prime*}A_{\nu}^\prime+\cdots\nonumber\\
&=&-\frac{N^2}{32{\pi}^2R}\int\frac{\rmd^4q}{(2\pi)^4}\sqrt{-g}g^{\mu\nu}g^{uu}A_{\mu}^{*}A_{\nu}^\prime\bigg|^{u=1}_{u=0}+\cdots.
\end{eqnarray}
In the last line, only surface term survives due to the equation of motion. Since the contribution from the horizon needs to be zero \cite{l}, one can write the action as
\begin{equation}
\fl S=\frac{N^2}{32\pi^2R}\int\frac{\rmd^4q}{(2\pi)^4}\sqrt{-g} g^{\mu\nu}g^{uu}f^*(q,u)A^{*}_{\mu}(q,u)A_{\nu}(q,u)\partial_u f(q,u)|_{u=0}.
\end{equation}
According to the minkowski space prescription \eref{pgreen}, we obtain
\begin{equation}
G_{\mu\nu}{A_{\nu}}|_{u\rightarrow 0}=-\frac{N^2}{16\pi^2R}\sqrt{-g}\ g^{\mu\nu}g^{uu}{A_{\nu}^\prime}|_{u\rightarrow 0}.
\end{equation}
We set the boundary conditions as
\begin{equation}
\lim\limits_{u\rightarrow 0}A_t=\alpha ,\qquad \lim\limits_{u\rightarrow 0}A_z=\beta,\qquad \lim\limits_{u\rightarrow 0}A_\alpha=\gamma,
\end{equation}
$A_t$ and $A_z$ are coupled, so we have the relation
\begin{equation}
G_{tt}\alpha+G_{tz}\beta=\frac{N^2T^2}{8}{A_{t}^\prime}\big|_{u=\epsilon},
\end{equation}
set $\beta=0$, we get $\alpha={A^{\prime\prime}_t uf}/{{\tilde{q}}^2}|_{u\rightarrow0}$ from \eref{motionb}, then the expression of $G_{tt}$ is
\begin{eqnarray}
G_{tt}=\frac{N^2T^2}{8}\frac{\tilde{q}^2A_t^\prime}{ufA^{\prime\prime}_t}\bigg|_{u=\epsilon}\label{gfunctiontt},
\end{eqnarray}
similarly, one obtains other components,
\begin{eqnarray}
G_{tz}=\frac{N^2T^2}{8}\frac{\tilde{q}\tilde{\omega}A_t^\prime}{ufA^{\prime\prime}_t}\bigg|_{u=\epsilon},\label{gfunctiontz}\\
G_{zz}=\frac{N^2T^2}{8}\frac{\tilde{\omega}^2A_t^\prime}{ufA^{\prime\prime}_t}\bigg|_{u=\epsilon},\label{gfunctionzz}\\
G_{\alpha\alpha}=-\frac{N^2T^2}{8}\frac{A^\prime_\alpha}{A_{\alpha}}\bigg|_{u=\epsilon},\label{gfunctionxx}
\end{eqnarray}
where $\epsilon$ is the infinitesimal which is relevant to the UV violation and can be handled by regularization.

\section{Thermal R-current correlators}\label{nresults}
In this section, the main task is to solve the equations \eref{motiona}-\eref{motiond}. \Eref{motionA} is a Fuchs equation to $A^\prime_t$ with four poles ($u=0,\pm 1,\infty$). What we need to know is the solution near $u=0$. Our numerical method starts from $u=1$ where we single out the ``incoming wave" \cite{l,p}. Then we run a differential equation program from $u=1$ all the way to $u=0$. The solution at $u=0$ suffers UV divergence, i.e. there is a logarithm term at $u\rightarrow 0$. To subtract this divergence, by renormalization, we must calculate the coefficient of the logarithm. Fortunately, one can obtain this coefficient by looking into the series solutions at both $u=0$ and $u=1$.

The two indices at $u=0$ are equal to zero. Then the two independent series solutions are
\numparts
\begin{eqnarray}
A_t^{\prime 1}&= [1+(\tilde{q}^2-\tilde{\omega}^2)u+\frac{1}{4}({\tilde{q}}^2-\tilde\omega^2)^2u^2]+\cdots,\\
A_t^{\prime 2}&=[1+(\tilde{q}^2-\tilde{ \omega}^2)u+\frac{1}{4}({\tilde{q}}^2-\tilde{ \omega}^2)^2u^2+\cdots]\ln u\nonumber\\
& +2(\tilde{ \omega}^2-\tilde{ q}^2)u+\frac{1}{4}[2-3(\tilde{ q}^2-\tilde{\omega}^2)^2]u^2+\cdots.
\end{eqnarray}
\endnumparts
So the solution of $A_t^\prime$ at $u=0$ is linear combination of $A_t^{\prime 1}$and $A_t^{\prime 2}$,
\begin{equation}\label{solutionu0}
A_t^\prime=a_1A_t^{\prime 1}+a_2A_t^{\prime 2}.
\end{equation}
Substitute the solution of $A_t^\prime$ \eref{solutionu0} into \eref{motionb} and take the limit $u\rightarrow 0$,  we get
\begin{equation}
a_2=(\tilde{q}^2 A_{t}+\tilde {\omega}\tilde{q}A_z)|_{u=\epsilon}=ufA^{\prime\prime}_t|_{u=\epsilon},
\end{equation}
where $\epsilon$ is a infinitesimal. This coefficient is exactly the coefficient of logarithm term of the solution of $A^{\prime}$ at $u=0$. This logarithm divergence originates from UV violation and can be regularized by general method in quantum field theory. We will subtract this term in our numerical result.

The indices at $u=1$ are $\nu_{\pm}=\pm{\rmi\tilde\omega}/{2}$. With the ``incoming wave'' boundary condition $\nu_{-}$, the corresponding series solution is
\begin{eqnarray}\label{solutionu1}
\fl A_t^\prime=(u-1)^{-\frac{\rmi\tilde{\omega}}{2}}\bigg[1+\frac{-2 \tilde{ q}^2 + 3 \rmi \tilde{\omega} +
 2 \tilde{\omega}^2 }{4(1-\tilde{\omega})}(u-1)\nonumber\\
\fl\qquad +\frac{24 \tilde{ q}^2 + 4 \tilde{ q}^4  -
 28 \rmi \tilde{  \omega}  - 24 \rmi \tilde {q}^2\tilde  {\omega } -
 42 \tilde{ \omega}^2 - 8 \tilde {q}^2 \tilde {\omega}^2  +
 23 \rmi\tilde {\omega}^3 + 4 \tilde {\omega}^4}{32 (\rmi + \tilde {\omega}) (2 \rmi + \tilde {\omega})}(u-1)^2\nonumber\\
\fl\qquad +\cdots\bigg].
\end{eqnarray}

\begin{figure}
  \begin{picture}(1.5in,1.5in)
    \put(2in,1.3in){(a)}
    \put(0in,0in){\includegraphics[width=2.5in,height=1.5in]{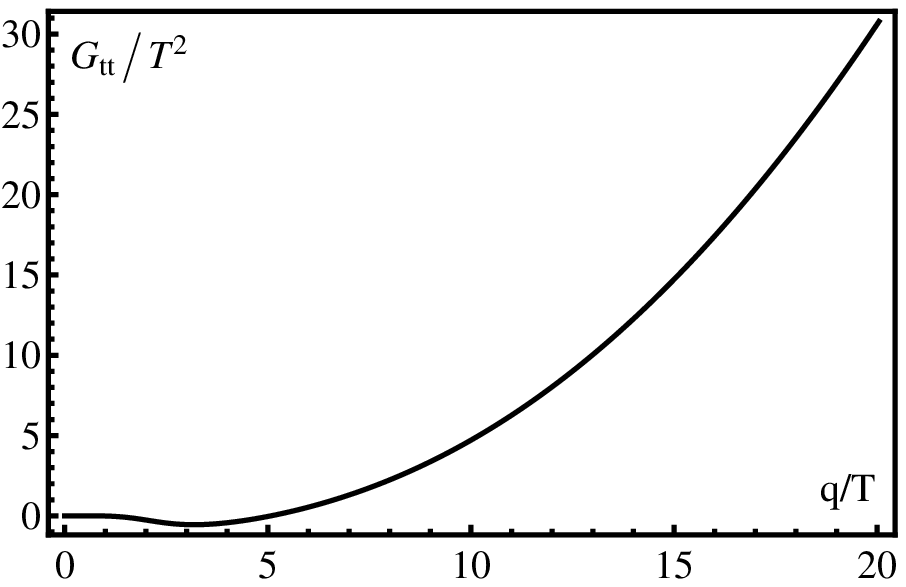}}
    \put(4.6in,1.3in){(b)}
    \put(2.6in,0in){\includegraphics[width=2.5in,height=1.5in]{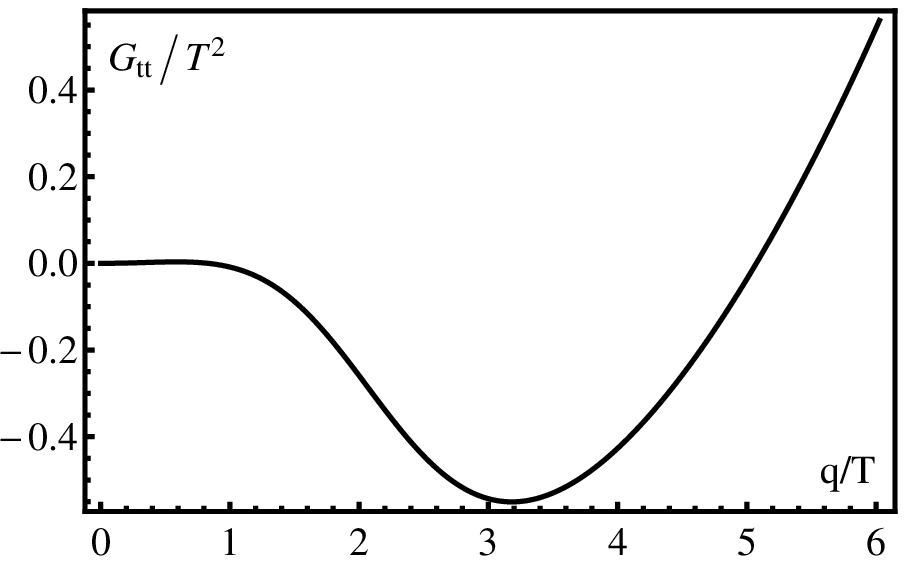}}
  \end{picture}
  \caption{\label{Gttq} The R-current correlator, $G_{tt}$, as a function of $q/T$ at the range of (a) $q/T\in[0, 20]$, (b) $q/T\in[0, 6]$ for $\omega=0.2$GeV, $T=0.2$GeV.}
\end{figure}

\begin{figure}
  \begin{picture}(1.5in,1.5in)
    \put(2in,1.2in){(a)}
    \put(0in,0in){\includegraphics[width=2.5in,height=1.5in]{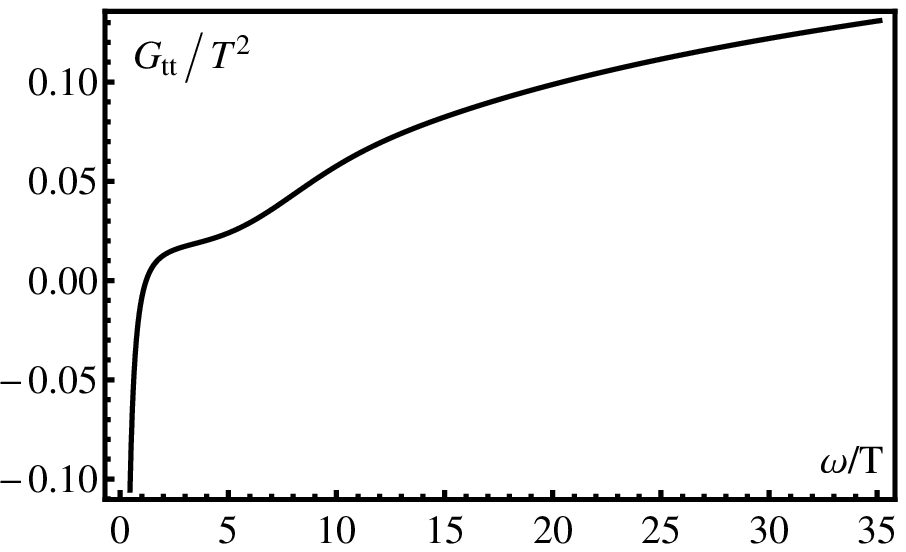}}
    \put(4.6in,1.2in){(b)}
    \put(2.6in,0in){\includegraphics[width=2.5in,height=1.5in]{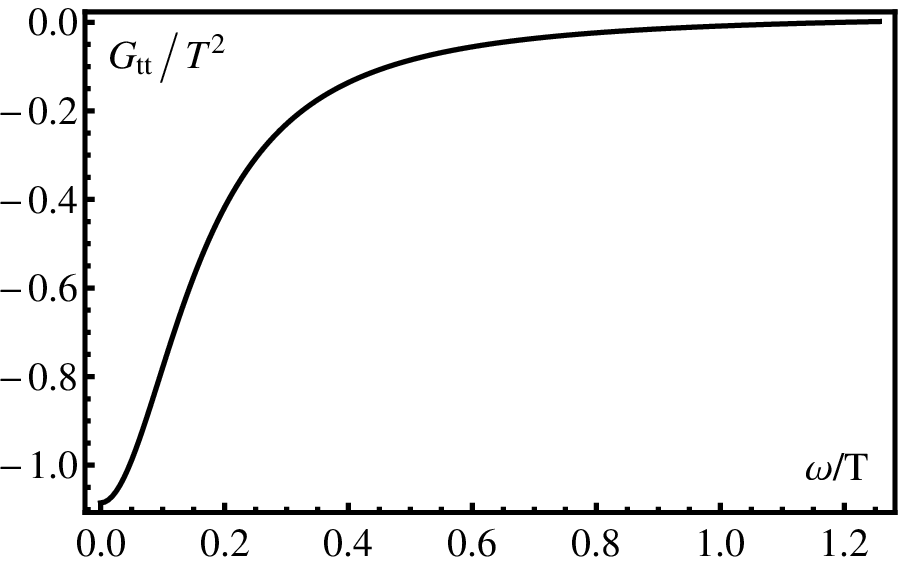}}
  \end{picture}
  \caption{\label{Gtto} The R-current correlator, $G_{tt}$, as a function of $\omega/T$ at the range of (a) $\omega/T\in[0, 35]$, (b) $q/T\in[0, 1.2]$ for $q=0.2$GeV, $T=0.2$GeV. }
\end{figure}

Take the value of \eref{solutionu1} at $u= 1-\epsilon$, as boundary condition of \eref{motionA}, one can run the program and obtain the value of $A^\prime_t$ at $u=\epsilon$. After subtracting the logarithm divergence, we finally obtain $G_{tt}$ in terms of $\omega$ and $q$ numerically. We show the results in \fref{Gttq} and \fref{Gtto}, in which we set the parameters as $N=3, T=0.2$GeV, $\epsilon=10^{-8}$ and the correlators are scalarized by $T^2$. \Fref{Gttq}(b) and \fref{Gtto}(b) are the details of \fref{Gttq}(a) and \fref{Gtto}(a) at small $q$ and $\omega$ respectively.

\begin{figure}
\begin{picture}(1.5in,1.5in)
    \put(1.5in,0in){\includegraphics[width=2.5in,height=1.5in]{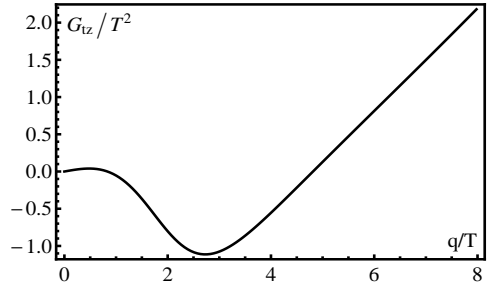}}
  \end{picture}
  \caption{\label{Gtzq}The R-current correlator, $G_{tz}$, as a function of q at the range of $q/T\in [0, 8]$ for $\omega=0.2$GeV, $T=0.2$GeV.}
\end{figure}

\begin{figure}
\begin{picture}(1.5in,1.5in)
    \put(1.5in,0in){\includegraphics[width=2.5in,height=1.5in]{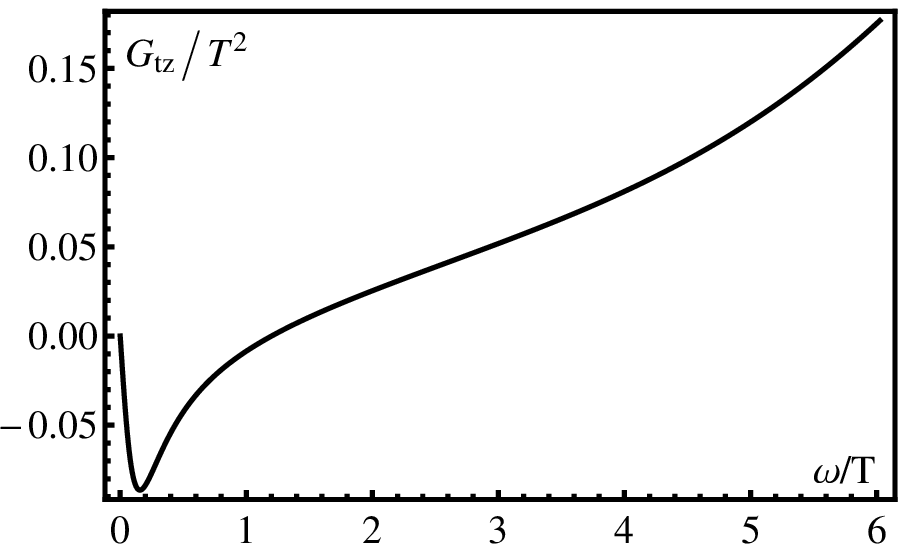}}
  \end{picture}
  \caption{\label{Gtzo}The R-current correlator, $G_{tz}$, as a function of $\omega$ at the range of $\omega/T\in [0, 6]$ for $q=0.2$GeV, $T=0.2$GeV.}
\end{figure}

\begin{figure}
\begin{picture}(1.5in,1.5in)
    \put(1.5in,0in){\includegraphics[width=2.5in,height=1.5in]{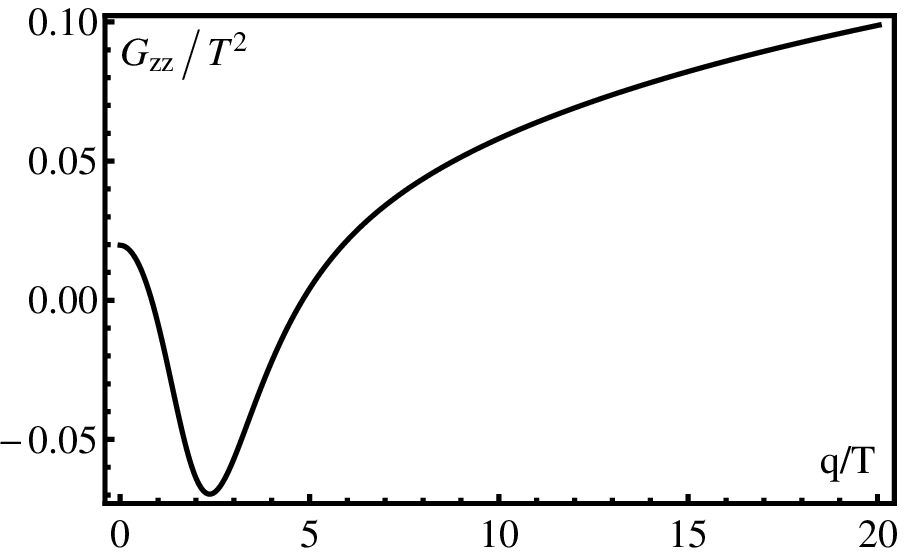}}
  \end{picture}
  \caption{\label{Gzzq} The R-current correlator, $G_{zz}$, as a function of $q/T$ at the range of $q/T\in[0, 20]$ for $\omega=0.2$GeV, $T=0.2$GeV. }
\end{figure}

\begin{figure}
\begin{picture}(1.5in,1.5in)
    \put(2in,1.3in){(a)}
    \put(0in,0in){\includegraphics[width=2.5in,height=1.5in]{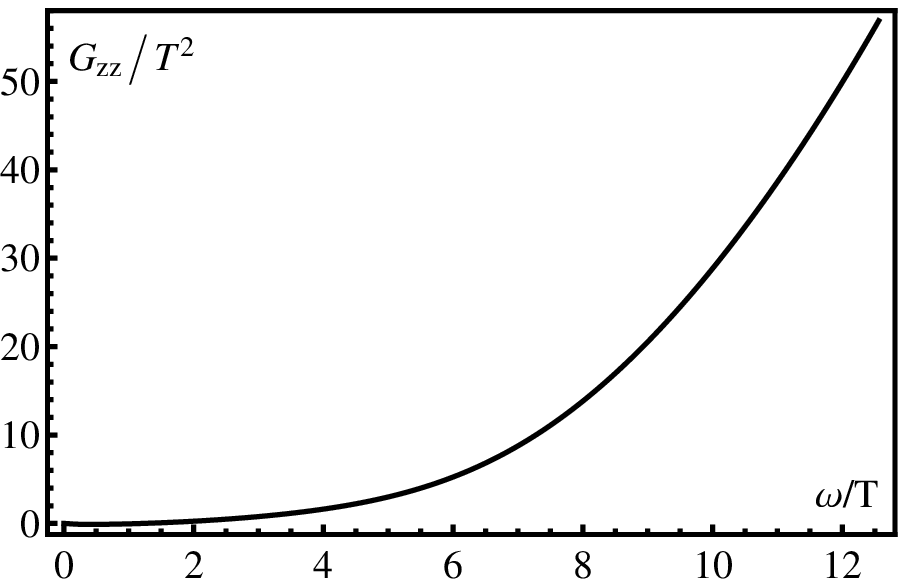}}
    \put(4.6in,1.3in){(b)}
    \put(2.6in,0in){\includegraphics[width=2.5in,height=1.5in]{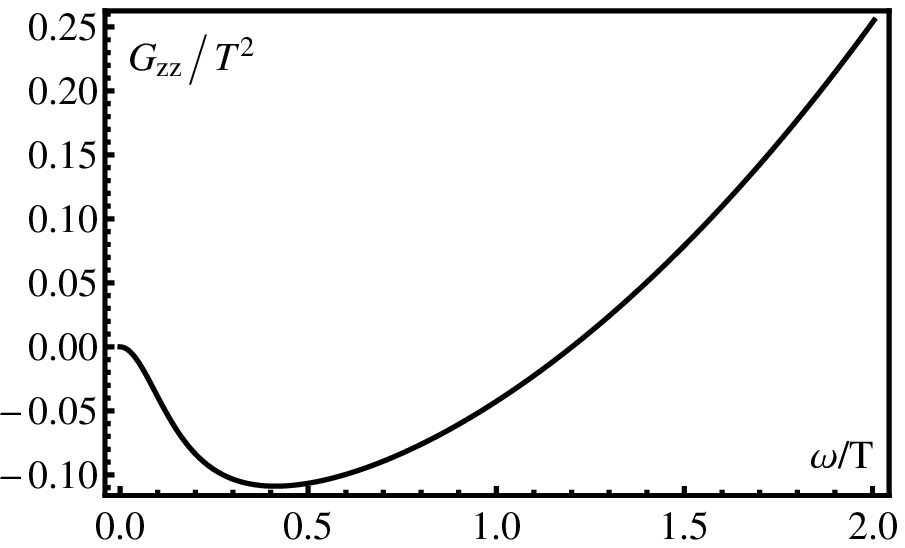}}
  \end{picture}
  \caption{\label{Gzzo} The R-current correlator, $G_{zz}$, as a function of $\omega/T$ at the range of (a) $\omega/T\in[0, 12]$, (b) $\omega/T\in[0, 2]$ for $q=0.2$GeV, $T=0.2$GeV. }
\end{figure}

We found that $G_{tt}/T^2$ is not a monotone increasing function of $q/T$. In \fref{Gttq}(b): there is an obvious minimum at $q/T\simeq 3.2$, and a local maximum, which is not easy to find, at $q/T\simeq 1$. The similar situation also emerge in other components, such as $G_{tz}/T^2$ varies with $q/T$ in \fref{Gtzq}, $G_{tz}/T^2$ varies with $\omega/T$ in \fref{Gtzo}, $G_{zz}/T^2$ varies with $q/T$ in \fref{Gzzq}, $G_{zz}/T^2$ varies with $\omega/T$ in \fref{Gzzo}.

Our results are in agreement with Plolicastro {\it et al}  \cite{m} at small $q$ and $\omega$($q/T,\omega/T\ll 1$). While the emerge of the extrema indicates that the correlators vary with energy and momentum in a rather complex way. All kinds of previous extreme results, say, the small $q$ and $\omega$ approximation is not sufficient to understand the complete properties of the correlators. So it is meaningful for us to investigate the correlators numerically in whole range of energy-momentum space.

Applying the same procedures, with the same parameters, we can also get numerical results of $G_{tz}$ and $G_{zz}$, as shown in figures \ref{Gtzq}-\ref{Gzzo}.

As for $A_\alpha$, the story is simpler since \eref{motiond} is decoupled for $G_{xx}$ and $G_{yy}$. It is also a four-poles ($0, \pm1,\infty$) Fuchs equation. The indices are 0 and 1 at $u=0$, which give rise to two independent series solution,
\begin{equation}
A_\alpha^{1}=\left[u+\frac{1}{2}(\tilde {q}^2-\tilde{\omega}^2)u^2\right]+\cdots\\
\end{equation}
and
\begin{equation}
A_\alpha^{2}=\left[u+\frac{1}{2}(\tilde {q}^2 -\tilde{\omega}^2)u^2+\cdots\right]\ln u+\frac{1}{\tilde{q}^2-\tilde{\omega}^2}+d_1u+d_2u^2\cdots.
\end{equation}
So that the series solution of $A_{\alpha}$ at $u=0$ is,
\begin{equation}
A_{\alpha}=a_1 A_{\alpha}^1+a_2 A_{\alpha}^2
\end{equation}
The coefficient of logarithm divergence $a_2$ can be obtained by taking the limit $u\rightarrow 0$ in \eref{motiond}, then the coefficient is computed as $a_2=uf(A^{\prime\prime}_{\alpha}+A^{\prime}_{\alpha}{f^{\prime}}/{f})|_{u=\epsilon}$.

At the pole of $u=1$, the indices are $\pm \rmi \tilde{\omega}/2$. We choose the ``incoming wave'' boundary condition at the horizon
\begin{eqnarray}
\fl A_\alpha =(u-1)^{-\frac{\rmi \tilde{\omega}}{2}}\bigg[1-(u-1) \frac{2 \rmi \tilde {q}^2  + \tilde{\omega} -
    2 \rmi \tilde{\omega}^2 }{
 4 (\rmi + \tilde{\omega})} \nonumber \\
\fl\qquad - (u-1)^{2}\frac{16 \tilde{ q}^2  + 4 \tilde{ q}^4 -
    4 \rmi \tilde{\omega} -
    16 \rmi \tilde {q}^2 \tilde{\omega} -
    18 \tilde{\omega}^2 -
    8 \tilde q^2 \tilde\omega^2  +
    15 \rmi \tilde{\omega}^3  + 4 \tilde{\omega}^4 }{
 32 (\rmi + \tilde{\omega}) (2 \rmi + \tilde{\omega})}\nonumber\\
 \fl\qquad+\cdots\bigg].
\end{eqnarray}

\begin{figure}
\begin{picture}(1.5in, 1.5in)
  \put(2in,1.3in){(a)}
  \put(0in,0in){\includegraphics[width=2.5in,height=1.5in]{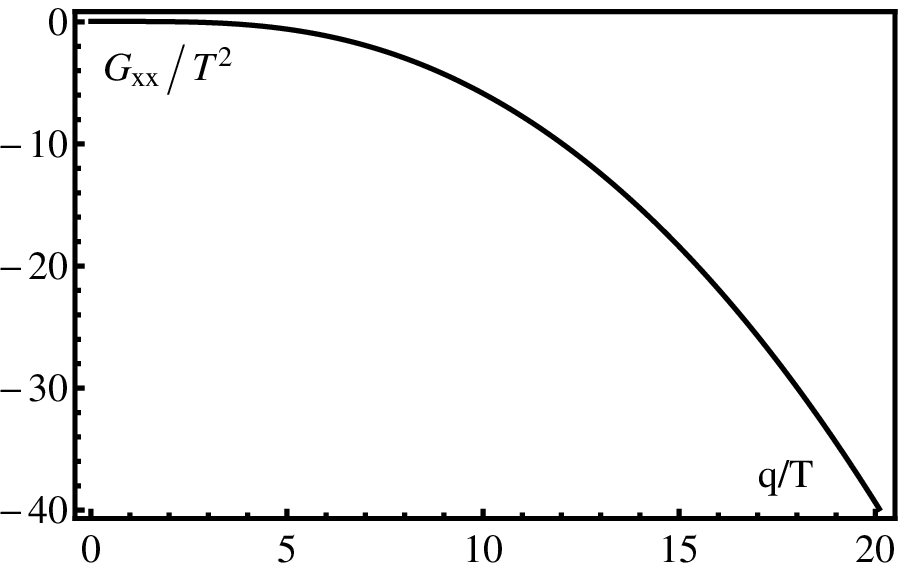}}
  \put(4.6in,1.3in){(b)}
  \put(2.6in,0in){\includegraphics[width=2.5in,height=1.5in]{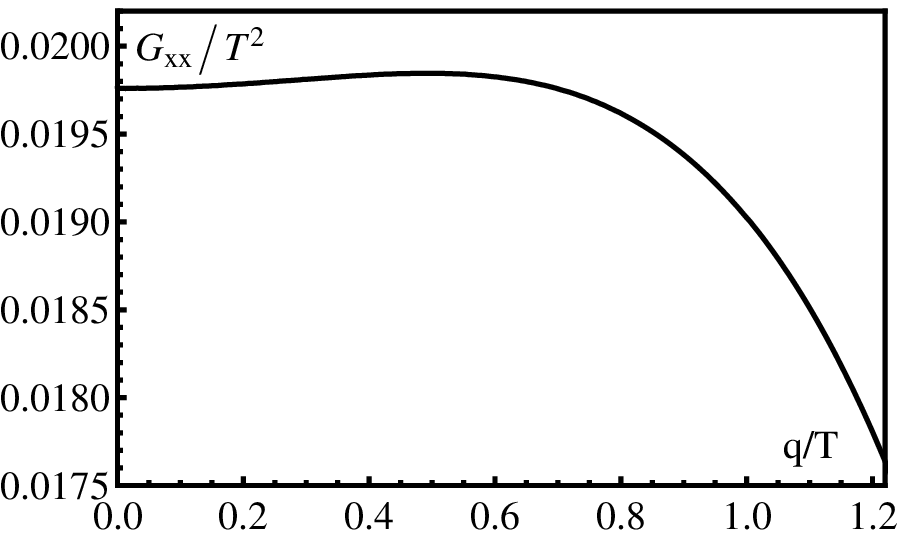}}
\end{picture}
  \caption{\label{Gxxq} The R-current correlator, $G_{\alpha\alpha}$, as a function of at the range of (a) $q/T\in[0, 20]$, (b) $q/T\in[0, 1.2]$ for $\omega=0.2$GeV, $T=0.2$GeV.}
\end{figure}

\begin{figure}
\begin{picture}(1.5in, 1.5in)
  \put(2in,0in){\includegraphics[width=2.5in,height=1.5in]{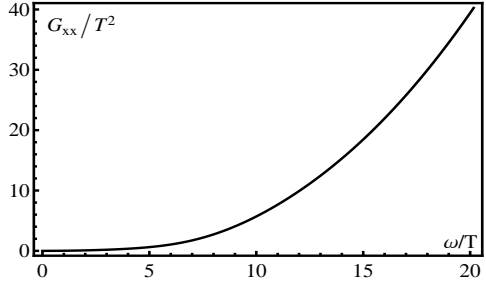}}
\end{picture}
  \caption{\label{Gxxo} The R-current correlator, $G_{\alpha\alpha}$, as a function of $\omega/T$ at the range of $\omega/T\in[0, 20]$ for $q=0.2$GeV, $T=0.2$GeV.}
\end{figure}

Similarly we set $N=3, T=0.2$GeV, and run a similar numerical program, and solve the equation of motion for $A_\alpha$. We have the curve of $G_{\alpha\alpha}/T^2$ varies with $q/T$ in \fref{Gxxq} at the range of (a) $q/T\in[0, 20]$, (b) $q/T\in[0, 1.2]$ for $\omega=0.2$GeV, and $G_{\alpha\alpha}$ varies with $\omega/T$ (\fref{Gxxo}) for $q=0.2$GeV. We also present $G_{\alpha\alpha}$ varies with $\omega/T$ at $q=0$ in \fref{Gi}(Dashed line), which is to compare with the analytical result for $G_{\alpha\alpha}$ in section \ref{aresults}.


\section{R-charge dielectric function}\label{dielectric}
From the field theory, we know that once we get the correlators, many actual physical quantities can be derived. As an application of the correlators, we discuss the dielectric function in this section.

In Coulomb gauge, the R-photon field dielectric function can be directly linked with self-energy $F(\omega,q)$ through \cite{q}
\begin{equation}\label{epsilon}
\varepsilon(\omega,q)=1-e^2\frac{F(\omega,q)}{q^2}.
\end{equation}

To compare our result with weak coupling, we will extract the T-dependent part of the correlators by subtracting $T=0$ contribution from $G_{tt}$.
When the temperature $T$ tends to $0$, we do the variable substitution $u=x T^2$ in \eref{motionA}, then we have the zero-temperature equation of motion,
\begin{eqnarray}
A_{t_0}^{\prime\prime\prime}+\frac{1}{x}A_{t_0}^{\prime\prime}+\frac{\omega^2-q^2}{(2\pi)^2x}A_{t_0}^{\prime}=0.\label{0t}
\end{eqnarray}
The finite solution of \eref{0t} is,
\begin{eqnarray}
{A^{\prime}_{t_0}}&=&K_0\left(\frac{\sqrt{x} \sqrt{q^2-\omega ^2}}{\pi }\right),\label{solutiont0u0}
\end{eqnarray}
where $K_{n}(z)$ are second kind modified Bessel functions.

\begin{figure}[!h]
\begin{picture}(1.5in,1.5in)
    \put(2in,1.3in){(a)}
    \put(0in,0in){\includegraphics[width=2.5in,height=1.55in]{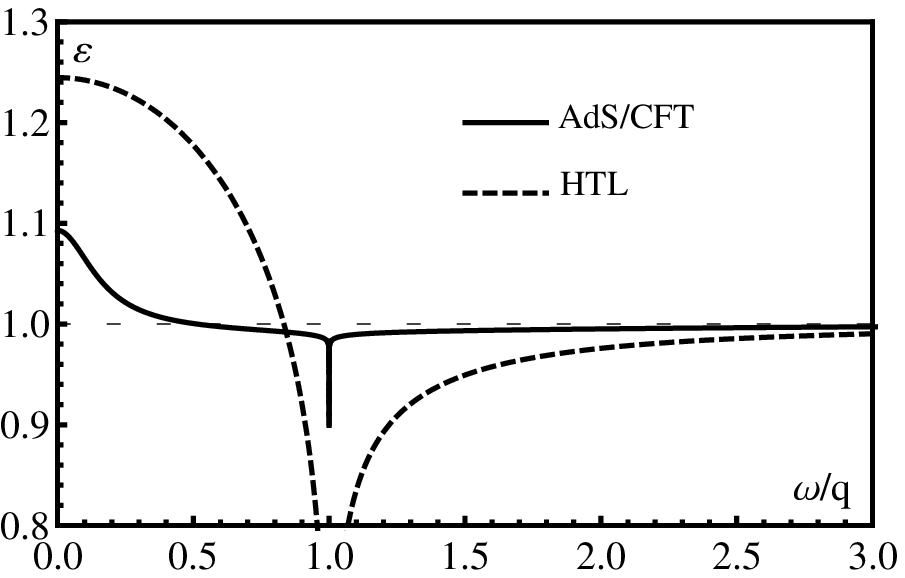}}
    \put(4.6in,1.3in){(b)}
    \put(2.6in,0in){\includegraphics[width=2.5in,height=1.5in]{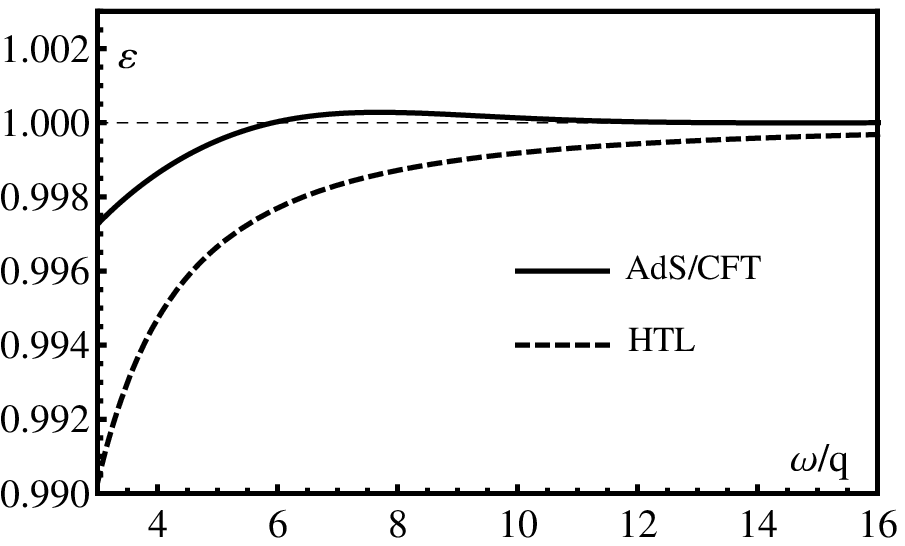}}
  \end{picture}
  \caption{\label{ep} R-photon dielectric function $\varepsilon(\omega, q)$ varies with $\omega/q$ at the range of (a) $\omega/q\in[0, 3]$ (b) $\omega/q\in[3, 16]$ for $q=0.2$GeV $T=0.2$GeV}
\end{figure}

Substitute the solution of $A^{\prime}_{t_0}$ \eref{solutiont0u0} into \eref{gfunctiontt}, then we get the finite temperature R-photon self-energy as
\begin{eqnarray}
F(\omega,q)=\frac{N^2}{8}\left[\frac{T^2{\tilde{q}}^2A^{\prime}_t}{ufA_{t}^{\prime\prime}}-\frac{q^2A_{t_0}^{\prime}}{(2\pi)^2xA_{t_0}^{\prime\prime}}\right]\Bigg|_{u\rightarrow 0},
\end{eqnarray}
where $x=u/T^2$. After setting $q=0.2$GeV, $T=0.2$GeV and $e=\sqrt{{4\pi}/{137}}$, we plot the dielectric function varies with $\omega/q$ in \fref{ep}, where (a) is the curve at the range of $\omega/q\in[0, 3]$, (b) is the details in large energy regime.

\begin{figure}[!h]
\begin{picture}(1.5in,1.5in)
    \put(0in,0.2in){\includegraphics[width=2.2in]{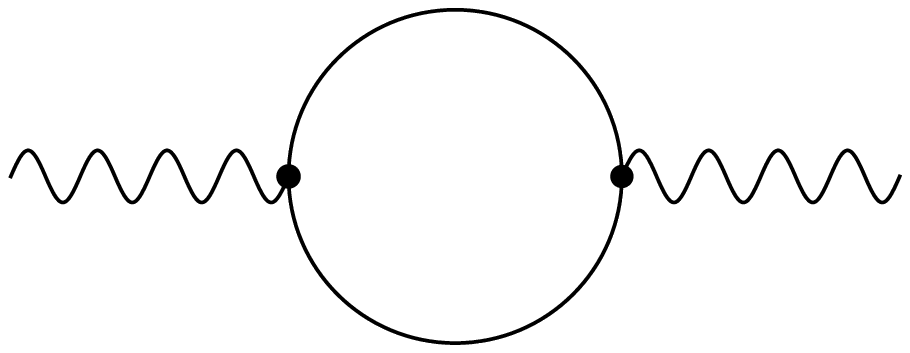}}
    \put(0.8in,0){(a)Fermi loop}
    \put(1.9in,0.2in){\includegraphics[width=2.2in]{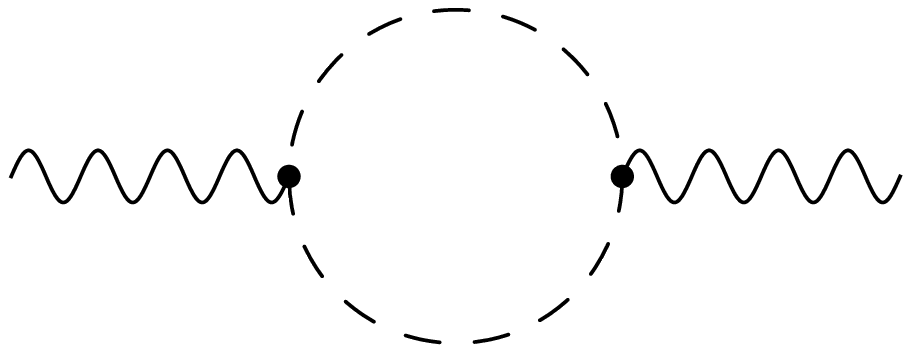}}
    \put(2.6in,0){(b)Scalar loop}
    \put(4.4in,0.2in){\includegraphics[width=1.1in]{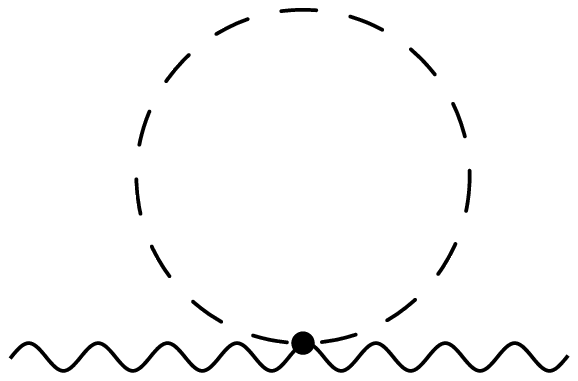}}
    \put(4.2in,0){(c)Scalar self-coupling}
\end{picture}
\caption{\label{selfenergy} One loop self-energy of R-photon. The solid line is for a Weyl fermi and the dashed line for a complex scalar.}
\end{figure}

At one-loop approximation, R-photon self-energy is determined by the three Feynman diagrams in \fref{selfenergy}.
In Coulomb gauge, we can calculate longitudinal retarded self-energy, which reads:
\begin{eqnarray}
\fl \ \ \ F_{L,R}^{a}(Q)=&\frac{(N^2-1)e^2}{2\pi^2}\int_{0}^{\infty}\rmd p\frac{p}{\rme^{\beta p}+1}\bigg[\frac{Q^2+4p^2+4\omega q}{4pq}\ln(\frac{Q^2+2\omega p+2pq+\rmi\epsilon}{Q^2+2\omega p-2pq+\rmi\epsilon})\nonumber \\
 &+\frac{Q^2+4p^2+4\omega q}{4pq}\ln(\frac{Q^2-2\omega p+2pq-\rmi\epsilon}{Q^2-2\omega p-2pq-\rmi\epsilon})-2\bigg]\\
\fl \ \ \ F_{L,R}^{b+c}(Q)=&\frac{(N^2-1)e^2}{8\pi^2}\int_{0}^{\infty}\rmd p\frac{p}{\rme^{\beta p}-1}\bigg[\frac{(\omega+2p)^2}{2pq}\ln(\frac{Q^2+2\omega p+2pq+\rmi\epsilon}{Q^2+2\omega p-2pq+\rmi\epsilon})\nonumber \\
 &+\frac{(\omega-2p)^2}{2pq}\ln(\frac{Q^2-2\omega p+2pq-\rmi\epsilon}{Q^2-2\omega p-2pq-\rmi\epsilon})-4\bigg].
\end{eqnarray}
With hard thermal loops approximation, namely $Q\sim eT, p\sim T$, and we obtain:
\begin{eqnarray}
F_{HTL}^{\omega,q}&=&\frac{(N^2-1)e^2 T^2}{3}\bigg[\frac{\omega}{2q}\ln (\frac{\omega+q+\rmi \epsilon}{\omega-q+\rmi\epsilon })-1\bigg].
\end{eqnarray}

Substitute HTL results into \eref{epsilon}, we can plot the dielectric function varies with $\omega/q$, and compare it with the AdS/CFT result in \fref{ep}(a), where the solid lines are for AdS/CFT and the dashed lines are for HTL. One observes that they both contain a mass-shell singularity at $\omega/q=1$, which separates the time-like and space-like regime, and the strong coupling singularity is sharper. At static limit $\omega=0$, the strong coupling is smaller than weak coupling (they are about 1.09 and 1.24 respectively). In the space-like regime strong coupling is always smaller than weak coupling, but in the time-like regime the situation is just opposite. This different will lead to different physical properties in space and time-regime. In the time-like regime, although the two curves approach to the vacuum limit $\epsilon=1$, the asymptotic behavior are different. HTL result monotonously increases and approaches to one from behind. While the AdS/CFT result increases first and then decreases to approach to the vacuum limit from above as shows in \fref{ep}(b). We do not know how to understand this difference in physics.

\section{Analytic results of $G_{\alpha\alpha}$ at $q=0$}\label{aresults}
Although we have obtained the numerical results for $G_{\alpha\alpha}$, we also find an analytic solution at long wavelength limit, i.e. $q=0$. In this limit the equation of motion for $A_{\alpha}$ is a three-pole (0,$\pm 1$) Fuchs equation as
\begin{equation}
A^{\prime\prime}_{\alpha}+\frac{f^\prime}{f}A_{\alpha}^\prime+\frac{\tilde{\omega}^2}{uf^2} A_{\alpha}=0\label{e},
\end{equation}
where $\alpha=x$ or $y$. $(0,1),(-\rmi\tilde\omega/{2}$,$\rmi\tilde\omega/{2})$ and $(-{\tilde\omega}/{2},
{\tilde\omega}/{2})$ are the indices for poles 0 , 1, -1 respectively.
In order to transform the equation of motion for $A_{\alpha}$ to a standard hypergeometric equation, one can do the variable transformation,
\numparts
\begin{eqnarray}
x&=\frac{2u}{1+u},\\
A_{\alpha}&=(x-1)^{-\frac{\rmi\tilde\omega}{2}}y,\label{transformation}
\end{eqnarray}
\endnumparts
then the equation of motion becomes
\begin{equation}
x(1-x)\frac{\rmd^2y}{\rmd x^2}+[-(-\rmi \tilde\omega +1)x]\frac{\rmd y}{\rmd x}-\frac{\tilde\omega ^2}{2}y=0.\label{Gii}
\end{equation}
The solutions for the above hypergeometric equation are
\numparts
\begin{eqnarray}
y_1(x)&=x*F\left(1-\frac{\rmi \tilde\omega +\tilde\omega }{2},1+\frac{\tilde\omega -\rmi \tilde\omega }{2},2,x\right),\\
y_2(x)&=x \bigg\{F\left(1-\frac{\rmi \tilde\omega +\tilde\omega }{2},1+\frac{\tilde\omega -\rmi \tilde\omega }{2},2,x\right)\ln \left(x\right)\nonumber\\
&+\frac{1}{\Gamma \left(1-\frac{\rmi \tilde\omega +\tilde\omega }{2}\right)\Gamma \left(1+\frac{\tilde\omega -\rmi \tilde\omega }{2}\right)}\Gamma \left(-\frac{\rmi \tilde\omega +\tilde\omega }{2}\right)\Gamma \left(\frac{\tilde\omega -\rmi \tilde\omega }{2}\right)x^{-1}\nonumber\\
&+\sum _{s=0}^{\infty } \frac{\left(1-\frac{\rmi \tilde\omega +\tilde\omega }{2}\right)_s\left(1+\frac{\tilde\omega -\rmi \tilde\omega }{2}\right)_s}{s!(2)_s}x^s\bigg[\psi \bigg(1-\frac{\rmi \tilde\omega +\tilde\omega }{2}\nonumber\\
&+s\bigg)+\psi \left(1+\frac{\tilde\omega -\rmi \tilde\omega }{2}+s\right)-\psi (2+s)-\psi (1+s)\bigg]\bigg\},\label{y2}
\end{eqnarray}
\endnumparts
where
\begin{eqnarray}
\Gamma(x)=\int_0^\infty \rme^{-t}t^{x-1}\rmd t  (x>1),\nonumber\\
(x)_n=\frac{\Gamma(x+n)}{\Gamma(x)}  (n\geq 1),\nonumber\\
\psi(x)=\frac{\Gamma^{\prime}(x)}{\Gamma(x)},\nonumber\\
F(\alpha,\beta,\gamma,x)=\sum^\infty_{n=0}\frac{(\alpha)_n(\beta)_n}{n!(\gamma)_n}x^n.\nonumber
\end{eqnarray}
The solution $y_1(x)$ indicates the "out-going wave condition" on boundary which results in the advanced correlator, so  only $y_2(x)$ survives to guarantee the "incoming wave condition" and retarded correlator. Then one inserts (\ref{y2}) into (\ref{transformation}), tansforms  $x$ back to $u$ and obtains
\begin{eqnarray}
\fl A_{\alpha}=\left(\frac{2u}{1+u}-1\right)^{-\frac{\rmi*\tilde\omega }{2}}\left(\frac{2u}{1+u}\right)\bigg\{F\left(1-\frac{\rmi \tilde\omega +\tilde\omega }{2},1+\frac{\tilde\omega -\rmi \tilde\omega }{2},2,\frac{2u}{1+u}\right)\nonumber\\
\fl \ln\left(\frac{2u}{1+u}\right)+\frac{1}{\Gamma\left(1-\frac{\rmi \tilde\omega +\tilde\omega }{2}\right)*\Gamma\left(1+\frac{\tilde\omega -\rmi \tilde\omega }{2}\right)}\Gamma\left(-\frac{\rmi \tilde\omega +\tilde\omega }{2}\right)\Gamma\left(\frac{\tilde\omega -\rmi \tilde\omega }{2}\right)\left(\frac{2u}{1+u}\right)^{-1}\nonumber\\
\fl +\left(\frac{\Gamma^{\prime}\left(1-\frac{\rmi \tilde\omega +\tilde\omega }{2}\right)}{\Gamma\left(1-\frac{\rmi \tilde\omega +\tilde\omega }{2}\right)}+\frac{\Gamma^{\prime}\left(1+\frac{\tilde\omega -\rmi \tilde\omega }{2}\right)}{\Gamma\left(1+\frac{\tilde\omega -\rmi \tilde\omega }{2}\right)}-\frac{\Gamma^{\prime}(2)}{\Gamma(2)}-\frac{\Gamma^{\prime}(1)}{\Gamma(1)}\right)\bigg\}.
\end{eqnarray}

Taking the derivatives on u, we get the logarithmic divergence term $D_{\alpha}$ in $A_{\alpha}^\prime$, that is
\begin{eqnarray}
D_{\alpha}&=&\frac{2}{1+u}\left(\frac{2u}{1+u}-1\right)^{-\frac{\rmi*\tilde\omega }{2}}\nonumber\\
& &F\left(1-\frac{\rmi \tilde\omega +\tilde\omega }{2},1+\frac{\tilde\omega -\rmi \tilde\omega }{2},2,\frac{2u}{1+u}\right)\ln \left(u\right)
\end{eqnarray}

\begin{figure}[!h]
  \centering
  \includegraphics[width=2.5in]{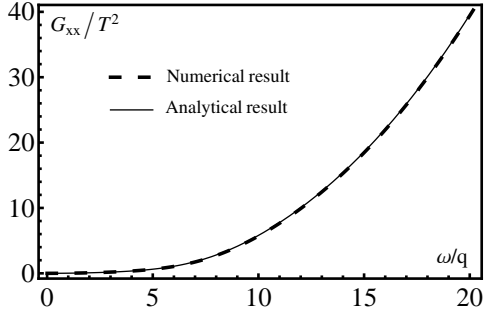}%
  \caption{\label{Gi} The R-current correlator, $G_{\alpha\alpha}$, as a function of $\omega/T$ at the range of $\omega/T\in[0, 20]$ for $q=0$, $T=0.2$GeV.}
\end{figure}

Applying the recipe, we get the solution of $G_{\alpha\alpha}$,
\begin{eqnarray}
 \fl G_{\alpha\alpha}=-\frac{N^2T^2}{8}\frac{A_{\alpha}^\prime-D_{\alpha}}{A_{\alpha}}\bigg|_{u\rightarrow 0}\nonumber\\
\fl=
\frac{N^2T}{16\pi}\omega  \left\{-\rmi+\left(\frac{\omega}{2\pi T}\right)  \left\{H\left[\left(\frac{1}{2}+\frac{\rmi}{2}\right)\frac{\omega}{2\pi T}\right] +H\left[\left(-\frac{1}{2}+\frac{\rmi}{2}\right)\frac{\omega}{2\pi T}\right]+\ln2\right\}\right\}.
\end{eqnarray}
where $H(a)$ is respect to Harmonic Number, $H(a )=\int_0^1 (1-x^{a })/(1-x) \, \rmd x$. The figure of $G_{\alpha\alpha}$ at $q=0$ and $T=0.2$GeV is shown in \fref{Gi}.

\section{Conclusions}\label{conclusions}
In this paper, by solving the equations of motion of bulk field in R-photon model, we calculated various components of Retarded correlators of R-current within the framework of AdS/CFT duality for non-zero momentum and energy.  Our numerical results, on one hand, are in agreement with \cite{m} which presented the analytic expressions for the correlators in $q\ll T$ and $\omega \ll T$ approximation, and on the other hand demonstrate some extremes of the correlators in terms of large energy and momentum  which were not discovered  in \cite{m}.  Besides the numerical results, we also found that the equation of motion for $A_x(A_y)$ has an analytic solution at $q=0$, which leads to an analytic expression for $G_{xx}(G_{yy})$.  The analytical expression is confirmed by our numerical result.

As an application of correlators, we presented the R-charge dielectric function  with respect to $\omega/q$ in strong coupling limit. The dielectric function shows a mass-shell singularity at $\omega/q=1$ which is also found  in weak coupling limit with HTL approximation. This singularity separates the time-like and space like regimes and results in different dielectric behaviors in these two regimes. In time-like regime, the strong coupling dielectric function approaches to vacuum limit from above, while weak coupling result from behind. The physical explanation for this difference is still absent.

\ack
We would like to thank Hai-cang Ren and Defu Hou for useful discussions and comments. This work is supported by NSFC under grant No. 10947002

\end{document}